\newcommand{\diracslash}[1]{#1\llap{/\kern2pt}}

\newcommand{\be}{\begin{equation}}
\newcommand{\ee}{\end{equation}}
\newcommand{\bea}{\begin{eqnarray}}
\newcommand{\eea}{\end{eqnarray}}
\newcommand{\ba}[1]{\begin{array}{#1}}
\newcommand{\ea}{\end{array}}

\documentclass[12pt]{iopart}
\usepackage{graphicx}
\begin{document}

\title{Finite temperature effects in light scattering off Cooper-paired Fermi atoms}

\author{Bimalendu Deb}

\address{ Department of Materials Science,
Indian Association for the Cultivation of Science, Jadavpur,
Kolkata - 700032.}
\begin{abstract}
We study stimulated light scattering off a superfluid  Fermi gas of
atoms at finite temperature.  We derive response function that takes
into account  vertex correction due to final state interactions; and
analyze finite temperature effects on collective and  quasiparticle
excitations of a uniform superfluid Fermi gas. Light polarization is
shown to play an important role in excitations. Our results suggest
that it is possible to excite Bogoliubov-Anderson phonon at a large
scattering length by light scattering.

\end{abstract}


\def\be{\begin{equation}}
\def\ee{\end{equation}}
\def\bearr{\begin{eqnarray}}
\def\eearr{\end{eqnarray}}
\def\zbf#1{{\bf {#1}}}
\def\bfm#1{\mbox{\boldmath $#1$}}
\def\hf{\frac{1}{2}}

\pacs{03.75.Ss,74.20.-z,32.80.Lg}

\maketitle

\section{Introduction}

Since the first realization of Bose-Einstein condensation in atomic
gases in 1995 \cite{bec}, there has been tremendous growth in
research activities with cold atoms. Recent experimental studies
\cite{hulet,solomon,thomas,mit,italy,grimm,expt} with cold fermionic
atoms have generated renewed interest in quantum many-body physics.
Atomic Fermi gases in traps provide a unique laboratory system for
exploring physics of interacting fermions with tunable interactions.
Fermi degeneracy in a trapped atomic  gas was first demonstrated
 by DeMacro and Jin
in 1999 \cite{jin}.  In recent past, there has been many reports of
possible observations of Fermi superfluidity (FS). However, an
unambiguous evidence of FS was shown by Zwierlein {\it et al.}
\cite{ketterle}. The  detection of pairing gap
\cite{gap1,gap2,theogap,zoller,huletp} and collective modes
\cite{duke,innsbruck,stringari} of FS are current issues of
interest. Physics at the crossover \cite{nozrink,randeria,crossover}
between BCS state of atoms and BEC of molecules \cite{molecules}
formed from Fermi atoms is of prime interest. A number of
theoretical investigations \cite{theory} have dealt with  FS near
crossover. Two very recent experiments \cite{huletim,ketterleim} on
two-component Fermi gases with imbalanced spin components provide
new insight on the nature of FS and perhaps indicate to some new
state of matter known as interior gap (IG) superfluidity predicted
by Liu and Wilczek \cite{liu}. The occurrence of IG superfluidity in
a two-component Fermi gas was theoretically predicted by Deb {\it et
al.} \cite{debig}.

In order to study the nature of FS, it is important to derive the
appropriate response function of Cooper-paired Fermi atoms due to an
external perturbation (such as photon or rf field). Our purpose here
is to calculate response function of superfluid Fermi gas at finite
temperature due to stimulated light scattering.  In a previous paper
\cite{deb}, we derived response function  at zero temperature. We
have also shown that it is possible to excite selectively a single
partner atom (of a particular hyperfine spin state) of a Cooper-pair
exploiting light polarizations in the presence of a strong magnetic
field. We present here detailed method of calculation of response
function at finite temperature due to light scattering. We study the
effects of finite temperature and light polarization  on
single-particle excitation  as well as collective mode of density
fluctuations. This collective mode known as Bogoliubov-Anderson (BA)
phonon \cite{bamode} has been theoretically studied in the context
of fermionic atoms \cite{mottelson}. We here show that it may  be
possible to excite this mode by light scattering: This mode appears
as a resonance in dynamic structure versus energy transfer in
scattering.  At finite temperature the resonance becomes broadened
due to Landau damping. In case of single-particle excitation
spectrum when Cooper-pairs are broken ($\omega > 2\Delta$), the peak
occurs at a higher energy as the temperature is lowered. If the
momentum transfer is higher, single-particle spectrum becomes
sharper. Light polarization significantly affects the  excitation
spectrum. Circular polarization of light leads to a positive shift
of the peak of single-particle spectrum at lower temperatures.

The paper is organized in the following way. In the following
section, we discuss in brief the stimulated light scattering off
Cooper-paired Fermi atoms. In Sec.II, we present in detail the
derivation of response function with vertex correction at finite
temperature. We next describe numerical results in Sec.III. The
paper is concluded in Sec.IV.

\section{Stimulated photon scattering}

Stimulated light scattering will occur when two laser beams with
nearly equal frequencies are impinged on atomic gas and tuned
far-off the resonance of a transition frequency of the atoms. Let us
specifically consider two-component Fermi gas of $^6$Li. Here the
two components imply the two hyperfine spin states of $F=1/2$ ground
level. As discussed previously \cite{deb}, large Zeeman shifts of
hyperfine sub-levels near Feshbach resonance allow us to utilize
light polarization to control  scattering to a significant extent.
Fig.1 of Ref. \cite{deb} describes polarization-selective light
scattering by which the amount of momentum and energy transfer to
the two partner atoms of a Cooper-pair can be controlled. It is
thereby possible to scatter photons from either spin component,
however because of long-range correlation between two components due
to  $s-$wave Cooper-pairing, the other spin component will also be
affected by photon scattering. When light fields are treated
classically, the effective atom-field interaction Hamiltonian is
\bearr H_{I} \propto \sum_{\sigma,\mathbf{k}} \gamma_{\sigma\sigma}
a_{\sigma,\mathbf{k}+\mathbf{q}}^{\dagger}a_{\sigma,\mathbf{k}}
\eearr where
$a_{\sigma,\mathbf{k}}(a_{\sigma,\mathbf{k}}^{\dagger})$ represents
the annihilation(creation) operator of an atom with hyperfine spin
$\sigma$ and center-of-mass momentum $\mathbf{k}$, $\mathbf{q}$ is
the momentum transfer and $\gamma_{\sigma\sigma}$ is the bare vertex
corresponding scattering without any change in spin state $\sigma$.

To describe density-density correlation function, we
 define the density operators  by $\rho_q^{(0)} =
\sum_{\sigma,\mathbf{k}}
a_{\sigma,\mathbf{k}+\mathbf{q}}^{\dagger}a_{\sigma,\mathbf{k}}$
and \bearr
 \rho_q^{(\gamma)} = \sum_{k,\sigma}
 \gamma_{\sigma\sigma} a_{\sigma,\mathbf{k}+\mathbf{q}}^{\dagger}a_{\sigma,\mathbf{k}}
 \eearr

\section{Response function}

The response function due to density fluctuation  is  \bearr
\chi(\mathbf{q},\tau-\tau') = -\langle
T_{\tau}[\rho_q^{(\gamma)}(\tau)\rho_{-q}^{(\gamma)}(\tau')]\rangle,
\eearr   where  $\langle \cdots \rangle$ implies thermal averaging
and $T_{\tau}$ is the complex time $\tau$ ordering operator. By
generalized fluctuation-dissipation theorem, the dynamic structure
factor  is given by \bearr S(\mathbf{q},\omega) = -
\frac{1}{\pi}[1+n_B(\omega)]\rm{Im}[ \chi(\mathbf{q},z=\omega +
i\delta)],  \label{flucdiss}\eearr where $\chi(\mathbf{q},z)$
represents Fourier transform of $\chi(\mathbf{q},\tau)$

\subsection{Vertex equation}
Within the framework of Nambu-Gorkov formalism of superconductivity,
using  Pauli matrices $\tau_i$, the vertex function can be written
as
 \cite{schrieffer} \bearr \Gamma(k_+,k_-) =
\tilde{\gamma} -  \int \beta^{-1}\frac{d^3\mathbf{k}'}{(2\pi)^3}
\sum_n \tau_3 \mathbf{G}(k_+')
\Gamma(k_+',k_-')\mathbf{G}(k_-')\tau_3V(\mathbf{k},\mathbf{k}'),
\label{vertex} \eearr  where $k_{+} \equiv
\{\mathbf{k}+\mathbf{q}/2,i(\omega_n + \nu_m/2)\}$ and $k_{-}
\equiv \{\mathbf{k}-\mathbf{q}/2,i(\omega_n-\nu_m/2) \}$ with
$\omega_n/\hbar = (2n + 1)(\hbar\beta)^{-1}$ and $\nu_m/\hbar =
2m(\hbar\beta)^{-1}$ representing the fermionic and bosonic
Matsubara frequency, respectively.   The Green function can be
expressed in a matrix form as \bearr
 G(k) = - \frac{i\mu_n \tau_{0} + \xi_k\tau_3 +
 \Delta_k\tau_1}{\mu_n^2 + E_k^2}, \label{green}\eearr where $\mu_n/\hbar$
 is
 Matsubara frequency, $E_k =
 \sqrt{\xi_k^2 + \Delta_k^2}$ and $\xi_k = \epsilon_k-\mu$ with $\epsilon_k = \hbar^2 k^2/(2m)$. The bare
 vertex can be written in a matrix form
  \bearr \tilde{\gamma} =
\gamma_0 \tau_0 + \gamma_3 \tau_3, \eearr where $\gamma_0 =
[\gamma_{\uparrow\uparrow} - \gamma_{\downarrow\downarrow}]/2$ and
$\gamma_3 = [\gamma_{\uparrow\uparrow} +
\gamma_{\downarrow\downarrow}]/2$.

The response function is related to the vertex function by

\bearr \chi(\mathbf{q},\omega) = \int
\frac{d^3k}{(2\pi)^3}\beta^{-1}\sum_n
\rm{Tr}[\tilde{\gamma}\mathbf{G}(k_{+})\Gamma(k_+,k_-)\mathbf{G}(k_-)]
\eearr

 To solve the vertex equation, let us expand the vertex function
 in terms of Pauli matrices as
 \bearr
\Gamma(k_+,k_-) =
\sum_{i=0}^{3}\Gamma^{(i)}(\mathbf{k},\mathbf{q},i\nu_m)\tau_i.
 \label{expansion}\eearr
We  replace $V(\mathbf{k},\mathbf{k}')$ by an effective  mean field
potential $V_{eff}$. In the weak-coupling limit, this potential
reduces to the form  $V_{\rm{weak}} = g a_s$ where $g=4\pi\hbar^2/m$
and $a_s$ is  s-wave  scattering length. Since we are interested
only in Cooper-pairing regime, we consider attractive interaction
only and hence  $a_s$ is assumed to be negative.  On replacing
$V(\mathbf{k},\mathbf{k}')$ by $g a_s$, the gap $\Delta$ becomes
$k$-independent. The vertex function also becomes $k-$independent
but remains a function of $q$ and $i\nu_m$ only.

Using Eqs. (\ref{green}) and (\ref{expansion}) in Eq.
(\ref{vertex}), we can write
 \bearr
&&\Gamma(\mathbf{k},\mathbf{q},i\nu_m) = \tilde{\gamma}_k - g a_s
\int \frac{d^3\mathbf{k}'}{(2\pi)^3}
\nonumber \\
&\times& \beta^{-1}\sum_n \frac{1}{[(\omega_n+\nu_m/2)^2 +
E_{k_+'}^2][(\omega_n-\nu_m/2)^2 +
E_{k_-'}^2)]}\sum_{i=0}^{3}y_i\tau_i, \label{vertexnew}\eearr where
$\xi_k =  \hbar^2k^2/(2m)-\mu$ and
 \begin{eqnarray} y_0 &=&
[i(\omega_n+\nu_m/2)\xi_{k_-'} +
i(\omega_n-\nu_m/2)\xi_{k_+'}]\Gamma^{(3)} -i\Delta[\xi_{k_+'} -
\xi_{k_-'}]\Gamma^{(2)} \nonumber \\
&+& \Delta[i(\omega_n+\nu_m/2)+i(\omega_n-\nu_m/2)]\Gamma^{(1)}
\nonumber \\
&+& [\xi_{k_+'}\xi_{k_-'} - (\omega_n+\nu_m/2)(\omega_n-\nu_m/2) +
\Delta^2]\Gamma^{(0)},
\end{eqnarray}
 \bearr y_1 &=&
-i[i(\omega_n+\nu_m/2)\xi_{k_-'} -
i(\omega_n-\nu_m/2)\xi_{k_+'}]\Gamma^{(2)} -\Delta[\xi_{k_+'} +
\xi_{k_-'}]\Gamma^{(3)}  \nonumber \\ &-& \Delta
[i(\omega_n+\nu_m/2)+ i(\omega_n-\nu_m/2)]\Gamma^{(0)} \nonumber \\
&+& [\xi_{k_+'}\xi_{k_-'} + (\omega_n+\nu_m/2)(\omega_n-\nu_m/2)
-\Delta^2]\Gamma^{(1)},  \eearr
 \bearr y_2 &=& - i\Delta [i(\omega_n+\nu_m/2)
- i(\omega_n-\nu_m/2)]\Gamma^{(3)} -i\Delta[\xi_{k_+'} -
\xi_{k_-'}]\Gamma^{(0)} \nonumber \\
&-& i [\xi_{k_+'}i(\omega_n-\nu_m/2) - i(\omega_n+\nu_m/2)\xi_{k-'}
 ]\Gamma^{(1)} \nonumber \\ &+& [\xi_{k_+'}\xi_{k_-'} +(\omega_n+\nu_m/2)(\omega_n-\nu_m/2) +
\Delta^2]\Gamma^{(2)}, \eearr
 \bearr y_3 &=& \Delta[\xi_{k_+'}+ \xi_{k_-'}]\Gamma^{(1)}
 -i\Delta[i(\omega_n+\nu_m/2) -
i(\omega_n-\nu_m/2)]\Gamma^{(2)}  \nonumber
\\ &+&
 [\xi_{k_+'}\xi_{k_-'}
  - (\omega_n+\nu_m/2)\omega_n - \Delta^2]\Gamma^{(3)} \nonumber \\
  &+& [\xi_{k_+'}i(\omega_n-\nu_m/2) + i(\omega_n+\nu_m/2)\xi_{k-'}]\Gamma^{(0)}.
 \eearr

There are basically two types of Matsubara frequency sums: \bearr
I_1(\mathbf{k}',\mathbf{q}) &=& \beta^{-1}\sum_n \left
[\frac{\omega_n+\nu_m/2}{(\omega_n+\nu_m/2)^2 +
E_{k_+'}^2}\right]\left
[\frac{\omega_n-\nu_m/2}{(\omega_n-\nu_m/2)^2 + E_{k_-'}^2}\right ],
\eearr \bearr I_2(\mathbf{k}',\mathbf{q}) &=& \beta^{-1}\sum_n
\frac{1}{(\omega_n+\nu_m/2)^2 + E_{k_+'}^2}\times
\frac{1}{(\omega_n-\nu_m/2)^2 + E_{k_-'}^2}\eearr  Now, the term
like $\omega/[\omega^2 + E_k^2]$ can be written in a separable form
\bearr \frac{\omega}{\omega^2 + E_k^2} &=& -\frac{1}{2i}\left[
\frac{1}{i\omega + E_{k}} + \frac{1}{i\omega
 - E_{k}} \right ], \eearr while the term like $1/[\omega^2 +
 E_k^2]$ can be decomposed as
\bearr \frac{1}{\omega^2 + E_k^2} &=& \frac{1}{2E_k}\left[
\frac{1}{i\omega + E_{k}} - \frac{1}{i\omega
 - E_{k}} \right ], \eearr The terms which are odd in
 frequency will not contribute to the sum and so can be omitted.
 Using these decompositions and after some algebra as done in Appendix-A, we can express
 \bearr I_1(\mathbf{k}',\mathbf{q}) = -\frac{1}{4} [T_{11} + T_{12} + T_{21} + T_{22}]
 \eearr
 \bearr I_2(\mathbf{k}',\mathbf{q}) =
\frac{1}{4E_{k_+'}E_{k_-'}} [T_{11} - T_{12} - T_{21} + T_{22} ]
\eearr
 where,
 \bearr T_{11} =  \frac{ \tanh(\beta
E_{k_-'}/2)-\tanh(\beta E_{k_+'}/2) }{2(E_{k_+'}-E_{k_-'} + i\nu_m)}
\label{t11} \eearr \bearr T_{12} = - \frac{ \tanh(\beta E_{k_-'}/2)+
\tanh(\beta E_{k_+'}/2)}{2(E_{k_+'}+E_{k_-'} + i\nu_m)} \label{t12}
\eearr \bearr T_{21} = - \frac{ \tanh(\beta E_{k_+'}/2) +
\tanh(\beta E_{k_-'}/2)}{2(E_{k_+'}+E_{k_-'} - i\nu_m)}
\label{t21}\eearr \bearr T_{22} &=& \frac{ \tanh(\beta E_{k_-'}/2) -
\tanh(\beta E_{k_+'}/2)}{2(E_{k_+'}-E_{k_-'} - i\nu_m)}
\label{t22}\eearr

In what follows, we use as the unit of energy the Fermi energy
$\epsilon_F = \hbar^2k_F^2/(2m)$ and accordingly scale all the
quantities. We denote $\tilde{\Delta} = \Delta/\epsilon_F$,
$\tilde{\xi}_k = \xi_k/\epsilon_F$, $\tilde{k}=k/k_F$ and so on. Let
$x=\tilde{q} \tilde{k} \cos \theta$, where $\theta$ is the angle
between $\mathbf{k}$ and $\mathbf{q}$.
 For notational convenience, let
$\mathcal{E} = \tilde{\xi}_k + q^2/4$,  $E_1=E_{k_-} =
\sqrt{(\mathcal{E} - x)^2 + \tilde{\Delta}^2}$ and $E_2= E_{k_+} =
\sqrt{(\mathcal{E} + x)^2 + \tilde{\Delta}^2}$.  Having now
performed the Matsubara frequency sum, omitting the terms which are
odd in $x$,  we can express the vertex equation as \bearr
\Gamma(\mathbf{k},\mathbf{q},i\nu_m) &=& \tilde{\gamma} - \kappa_s
\int \frac{d^3\mathbf{k}'}{(2\pi)^3}
\sum_{i=0}^{3}c_i(\mathbf{k}')\tau_i, \label{verteqfinal}\eearr
where $\kappa_s = g a_s k_F^3/\epsilon_F $ and
\begin{eqnarray} c_0 &=&
[(\mathcal{E}^2 - x^2 + \tilde{\Delta}^2)I_2 - I_{1}]\Gamma^{(0)},
\nonumber
\end{eqnarray}
 \bearr c_1 &=& [\nu_m \Gamma^{(2)} -2\tilde{\Delta} \Gamma^{(3)}] \mathcal{E} I_2
 +
[(\mathcal{E}^2 - x^2-\tilde{\Delta}^2)I_2 + I_1]\Gamma^{(1)},
\nonumber \eearr
 \bearr c_2 &=&  \tilde{\Delta} \nu_m I_2\Gamma^{(3)} - \nu_m\mathcal{E}
I_2\Gamma^{(1)} + [(\mathcal{E}^2 - x^2 + \tilde{\Delta}^2)I_2 +
I_1]\Gamma^{(2)}, \nonumber \eearr
 \bearr c_3 &=&  2\tilde{\Delta}\mathcal{E}I_2\Gamma^{(1)}
 + \tilde{\Delta} \nu_m I_2\Gamma^{(2)} +
 [(\mathcal{E}^2 - x^2 - \tilde{\Delta}^2)I_2
-I_1 ]\Gamma^{(3)}
 \nonumber
 \eearr
Now, the vertex terms $\Gamma^{(i)}$ form four coupled algebraic
equations \bearr \Gamma^{(i)} = \gamma_i - \kappa_s \int
\frac{d^3\tilde{\mathbf{k}}'}{(2\pi)^3} c_i, \hspace{0.5cm} i=0,3
\label{vert03} \eearr \bearr \Gamma^{(j)} =  - \kappa_s \int
\frac{d^3\tilde{\mathbf{k}}'}{(2\pi)^3} c_j. \hspace{0.5cm} j=1,2
\label{ver12} \eearr In the limit $q \rightarrow 0$, $i\nu_m
\rightarrow 0$ and $\Gamma^{(2)} \rightarrow \tilde{\Delta}$, the
equation for $\Gamma^{(2)}$ reduces to the standard BCS gap equation
\bearr \tilde{\Delta} = -\kappa_s \int
\frac{d^3\tilde{\mathbf{k}}}{(2\pi)^3} \frac{\tanh(\beta
E_k/2)}{2E_{k}}\tilde{\Delta} \label{gap} \eearr which implies
$|\kappa_s| I_0 = 1$ where \bearr I_0 = \int
\frac{d^3\tilde{\mathbf{k}}}{(2\pi)^3} \frac{\tanh(\beta
E_k/2)}{2E_{k}}. \label{i0} \eearr The gap equation expressed in
this form has logarithmic divergence. However, this divergence can
be removed by renormalizing the mean-field interaction via
subtracting the zero temperature and zero pairing-field ($\Delta=0$)
part of the integral. Although this gap equation resembles the
standard weak-coupling BCS gap equation, the chemical potential
$\mu$ can deviate significantly from its weak-coupling value $\mu
\simeq \epsilon_F$. The chemical potential is given by single-spin
superfluid number density  \bearr n= \frac{1}{2} \int \frac{
d^3\mathbf{k}}{(2\pi)^3}\left (1-
\frac{\tilde{\xi}_k}{E_k}\tanh(\beta E_k /2)\right ). \label{num}
\eearr

Let us first consider the off-diagonal terms $\Gamma^{(1)}$ and
$\Gamma^{(2)}$. Making use of the relation (\ref{i0}) and the
analytic continuation $i\nu_m \rightarrow \omega + i0^{+}$, \bearr
\Gamma^{(1)} = \frac{i\tilde{\omega} J_1 \Gamma^{(2)} +
2\tilde{\Delta} J_1 \Gamma^{(3)}}{ (\tilde{I}_2 - \tilde{\Delta}^2
J_2 + \tilde{I}_1) - I_0 } \label{g1} \eearr \bearr \Gamma^{(2)} =
\frac{i\tilde{\Delta} \tilde{\omega} J_2\Gamma^{(3)} -
i\tilde{\omega} J_1 \Gamma^{(1)}}{(\tilde{I}_2 + \tilde{\Delta}^2
J_2 + \tilde{I}_1) - I_0 } \label{g2} \eearr where the various
integrals are as follows \bearr \tilde{I}_1(q,i\nu_m\rightarrow
\omega + i0^+) = \int
\frac{d^3\mathbf{k}}{(2\pi)^3}I_1(\mathbf{k},\mathbf{q}) \eearr
\bearr J_1(\mathbf{q},i\nu_m \rightarrow \tilde{\omega} + i0^+) &=&
\int \frac{d^3\mathbf{k}}{(2\pi)^3}\mathcal{E}
I_2(\mathbf{k},\mathbf{q}) \eearr \bearr J_2(\mathbf{q},i\nu_m
\rightarrow \omega + i0^+)&=& \int
\frac{d^3\mathbf{k}}{(2\pi)^3}I_2(\mathbf{k},\mathbf{q}) \eearr
\bearr \tilde{I}_2(\mathbf{q},i\nu_m \rightarrow \omega + i0^+) =
\int \frac{d^3\mathbf{k}}{(2\pi)^3}(\mathcal{E}^2 - x^2)
I_2(\mathbf{k},\mathbf{q}). \eearr  The method of calculation of
various integrals is described in Appendix-A. Eliminating
$\Gamma^{(1)}$ from Eqs. (\ref{g1}) and (\ref{g2}), we have \bearr
\Gamma^{(2)} = i\tilde{\Delta}\tilde{\omega} \left
[1-\frac{(\tilde{\omega} J_1)^2}{D_{+}D_-}\right ]^{-1}\left [
\frac{ J_2}{D_+} + \frac{2 J_1^2}{D_+D_-} \right ] \Gamma^{(3)}
\eearr where $D_{\pm} = (\tilde{I}_1 + \tilde{I}_2-I_0) \pm
\tilde{\Delta}^2 J_2$. Here we note  that although the integral
$\tilde{I}_1$, $\tilde{I}_2$ and $I_0$ have logarithmic divergence,
this divergence does not pose any problem. Because, at the end all
the divergences are exactly canceled out and thus  all the vertex
terms $\Gamma^{(i)}$'s and the response function remain finite. The
integrals $J_1$ and $J_2$ are finite.  From Eq. (\ref{vert03}), we
have the solution \bearr \Gamma^{(0)} = \frac{\gamma_0}{1+\kappa_s
B} \eearr where \bearr B = \tilde{I}_2 + \tilde{\Delta}^2 J_2 -
\tilde{I}_1. \eearr On substitution of Eqs. (\ref{g1}) and
(\ref{g2}), we have the solution \bearr \Gamma^{(3)} =
\frac{\gamma_3}{1+\kappa_s F} \label{g3} \eearr where \bearr F = &&
A + \left [(\tilde{\Delta} \tilde{\omega})^2 J_2 -
\frac{2(\tilde{\Delta} \tilde{\omega} J_1)^2}{D_-}\right ] \left [
\frac{
J_2}{D_+} + \frac{2 J_1^2}{D_+D_-} \right ] \nonumber \\
&\times& \left [1-\frac{(\tilde{\omega} J_1)^2}{D_{+}D_-}\right
]^{-1} + \frac{(2\tilde{\Delta} J_1)^2}{D_-} \eearr with \bearr A =
\bar{I}_2 - \tilde{\Delta}^2 J_2 - \bar{I}_1 \eearr

After having done Matsubara frequency sum and analytic continuation,
we can write the response function as [see Appendix-B] \bearr
\chi(q,\omega) = 2 B \gamma_0\Gamma^{(0)}+ 2 A\gamma_3\Gamma^{(3)} -
2 i \tilde{\omega} \tilde{\Delta} J_2 \gamma_3 \Gamma^{(2)}
+4\tilde{\Delta} J_1 \gamma_3\Gamma^{(1)} \eearr On replacing
$\Gamma^{(2)}$ and $\Gamma^{(1)}$ in terms of $\Gamma^{(3)}$, we
have \bearr \chi(q,\omega) &=& 2 B
\gamma_0\Gamma^{(0)} + 2 F \gamma_3\Gamma^{(3)} \nonumber \\
&=& \frac{2B}{1+\kappa_s B} \gamma_0^2 + \frac{2F}{1+\kappa_s F}
\gamma_3^2 \eearr Now, the dynamic structure factor can easily be
written as \bearr S(\mathbf{q},\omega) &=& -
\frac{1}{\pi}[1+n_B(\omega)]\rm{Im}[ \chi(q,\omega)].
\label{flucdiss} \nonumber \\
&=& - \frac{2}{\pi}[1+n_B(\omega)] \left [
\frac{\gamma_0^2\rm{Im}(B)}{|1+\kappa_sB|^2} + \frac{
\gamma_3^2\rm{Im}(F)}{|1+\kappa_sF|^2} \right ] \label{chi} \eearr

\section{results and discussions}

\begin{figure}
 \includegraphics[width=4.5in,height=3.0in]{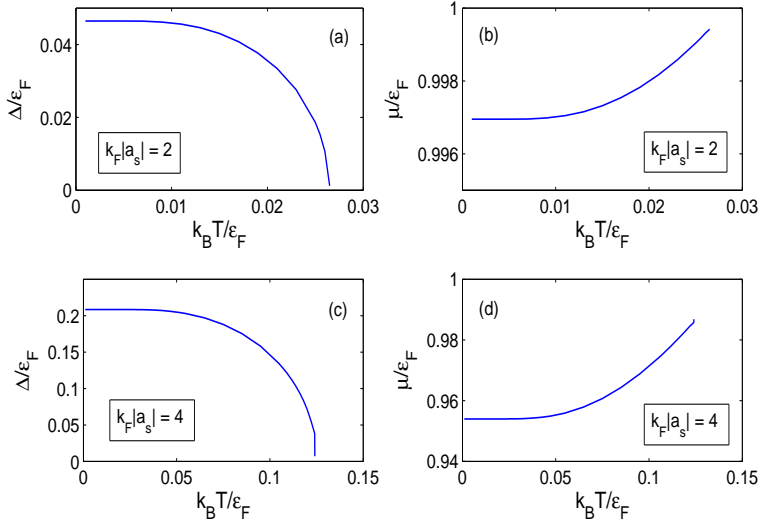}
 \caption{Gap $\Delta$ (in unit of $\epsilon_F$) is plotted as a function of temperature
 $k_B T$ (in unit of $\epsilon_F$) for $k_F |a_s| = 2$ (a) and $k_F |a_s| = 4$ (c), while
 subplots (b) and (d) exhibit the variation of
 the chemical potential $\mu$ as a function of temperature for $k_F |a_s| = 2$ and $k_F |a_s| = 4$,
 respectively. }
 \label{fig1}
 \end{figure}
 Before we elaborate our results, we note a  few
pertinent points. First, in calculating response function, unlike
weak-coupling case,  all the energy integrations are carried out
over the entire energy range meaning that $\xi$ ranges from $-\mu$
to $\infty$.  Second, as regards vertex correction in light
scattering, our formalism of calculating response function can
account for any arbitrary polarization of incident light. Third, we
have devised a procedure whereby we carry out first the angular
integrations by parts considering $\omega$ as a complex parameter
$z$ as described in  Appendix-A. This leads to trigonometric
functions of $z$,  which are then easily analytically continued
using the limit $z \rightarrow \omega + i 0^{+}$. This is unlike the
usual approach where the functions like $1/[\omega \pm (E_{\pm})  +
i 0^{+}]$ are first separated into real and imaginary parts. The
imaginary part is a delta function, the energy and angular
integrations over which require finding the roots of a complicated
polynomial equation.

\begin{figure}
 \includegraphics[width=4.5in,height=3.0in]{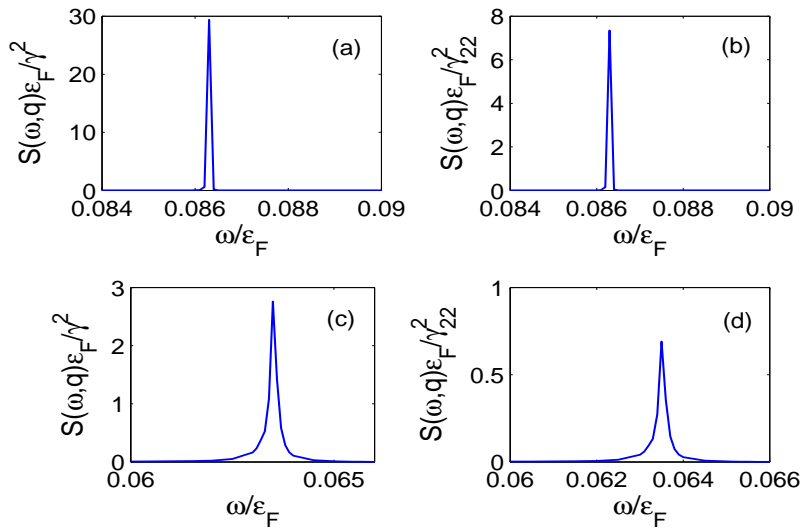}
 \caption{Subplot-(a): Dynamic structure factor $S(\omega,q)$ (in dimensionless unit) is plotted as a function of
 energy transfer
 $\omega$ (in unit of $\epsilon_F$) for $q = 0.3 k_F$, $k_F|a_s| = 2.0$, $k_B T =
 0.009 \epsilon_F$ ($T_c = 0.0265 \epsilon_F$) for unpolarized light with $\gamma_{11} = \gamma_{22} =
 \gamma$, that is, $\gamma_0 = 0$ and $\gamma_3 = \gamma$.
 Subplot-b: Same as in (a), but for circularly polarized light with
 $\gamma_{11} =0$ and $\gamma_{22} \ne 0$, that is, $|\gamma_0| =
 |\gamma_3| = \gamma_{22}/2$. Subplots (c) and (d) are the counterparts of (a) and (b), respectively;
 but for $k_B T = 0.021$.}
 \label{fig2}
 \end{figure}

 For all our numerical illustrations, the coupling
strength $\kappa_s$ is taken to be negative since we are interested
in Cooper-pairing regime only. Figure 1 shows temperature variation
of $\Delta$ and $\mu$ for two different scattering lengths $a_s$.
Comparing Fig.1 (a) with Fig.1 (c), we note that  critical
temperature becomes large for large scattering length.  This implies
that finite temperature effects are particularly important for large
scattering lengths. In the zero temperature limit, the gap becomes
almost independent of temperature, but remain largely dependent on
interaction parameter. As in our previous paper \cite{deb} for
$T=0$, for numerical illustration we consider two cases: Case-I:
When both the laser beams  are unpolarized, that is, $\gamma_{11}
(\equiv \gamma_{\uparrow\uparrow}) = \gamma_{22} (\equiv
\gamma_{\downarrow\downarrow})$; Case-II: When both the beams are
either $\sigma_+$ or $\sigma_-$ polarized. These are two limiting
cases only, we reemphasize that our formalism can handle any
arbitrary polarization.

\begin{figure}
 \includegraphics[width=4.5in,height=3.0in]{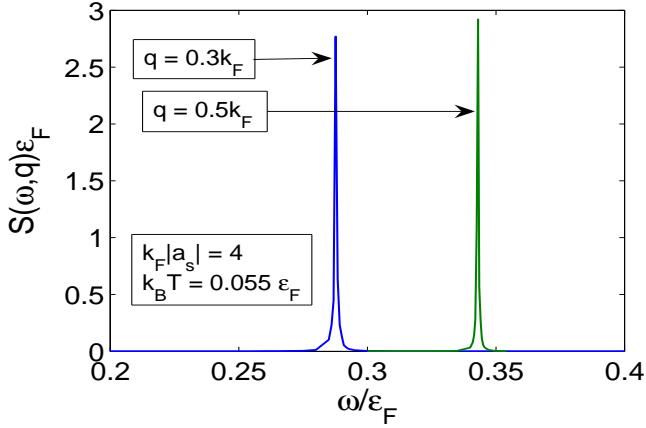}
 \caption{ $S(\omega,q)$  as a function of
 $\omega$ for two different values of $q$ as shown in the figure. The other parameters are
  $k_F|a_s| = 4.0$, $k_B T =
 0.055 \epsilon_F$ ($T_c = 0.124 \epsilon_F$, $\Delta = 0.203 \epsilon_F$), $\gamma_{11} = \gamma_{22} =
 1$.}
 \label{fig3}
 \end{figure}

\begin{figure}
 \includegraphics[width=4.5in,height=3.0in]{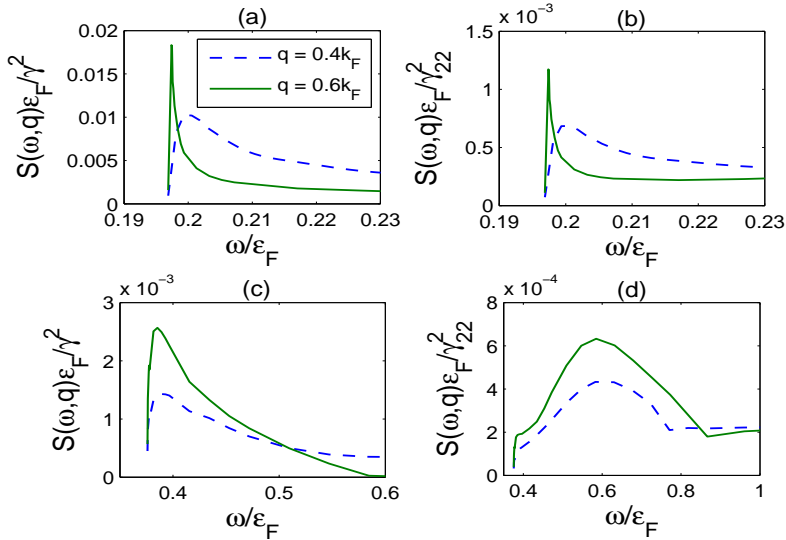}
 \caption{ Quasiparticle excitation spectrum ($\omega > 2\Delta$)
 for a uniform Fermi superfluid at  two different
 temperatures $k_BT = 0.115 \epsilon_F$ (a, b) and $k_BT = 0.075
 \epsilon_F$ (c, d), but for the same
  $k_F|a_s| = 4.0$. The subplots (a) and (c) are for unpolarized light
  with $\gamma_{11} = \gamma_{22} = \gamma$, while (b) and (d) are
  for circularly polarized light with $\gamma_{11} = 0$.
  The dashed curve refers to $q = 0.4 k_F$ and the solid one
  to $q = 0.6 k_F$ in all the subplots. $\Delta$ is 0.098$\epsilon_F$
  and 0.188$\epsilon_F$ at temperatures
  $0.115 \epsilon_F$ and $0.075 \epsilon_F$, respectively.}
 \label{fig4}
 \end{figure}

Let us next discuss the regime of collective excitation of
density-density fluctuations for energies $\omega < 2\Delta$. This
regime is characterized by long-wave mode of vibration known as
Bogoliubov-Anderson (BA) phonon which results from vertex
correction.  The general expression for $\chi$ as given by Eq.
(\ref{chi})  has two parts which are proportional to $\gamma_0^2$
and $\gamma_3^2$, respectively. Since the real part of $B$ is always
negative, the first part is always finite. When $\Gamma^{(0)}
\rightarrow \gamma_0$, $\Gamma^{(3)} \rightarrow \gamma_3$ and
$\Gamma^{i} \rightarrow 0$ with $i=1,2$, that is, when vertex
correction is neglected, the response function reduces to \bearr
\chi(q,\omega) = 2B \gamma_0^2 + 2A \gamma_3^2 \eearr which has no
pole. However, when vertex correction is added, there arises the
pole of $\chi$ which is given by \bearr
 F = \frac{1}{|\kappa_s|}. \label{ba} \eearr
In the limit $|a_s| \rightarrow 0$, the pole is determined by $ D_+
= 0$. The real part of the pole will correspond to the phonon energy
while the negative imaginary part will give the damping of the mode.
Only in the low $q$-regime and  $T < T_c$, BA mode will be well
defined. For larger temperatures (but $<T_c$), this mode will be ill
defined because of large  Landau damping which occurs due to its
coupling with thermally excited quasiparticles.  At zero
temperature, in the limit $a_s \rightarrow 0^{-}$, we have
reproduced the standard result \bearr \omega_{\rm{BA}} =
\frac{1}{\sqrt{3}} \hbar q v_F \eearr as calculated in Appendix-C.
The BA phonon appears as a resonance in the spectrum of dynamic
structure factor as is demonstrated in Figs. 2 and 3. Comparing
Fig.2(a) with Fig.2(c), we notice that as the temperature is
increased, the spectrum becomes broadened. The resonance energy does
not depend on the state of polarization of light. However,  the peak
and width of the resonance  may depend on light polarization. Figure
4 displays the quasiparticle excitation spectrum when $\omega$
exceeds the pair-breaking energy $2\Delta$. As temperature is
decreased, the peak makes a positive shift because of increase in
$\Delta$. For larger $q$ values, the peak is larger. At lower
temperatures, polarization of light significantly affect the
spectrum as can be noticed by comparing Fig. 4(c) with 4(d).

\section{conclusion}

In conclusion, we have derived finite temperature response function
of superfluid Fermi gas due to scattering of polarized light. The
response function takes into account vertex correction due to final
state interactions. We have presented selective results on dynamic
structure factor (DSF) deduced from the response function with the
aid of generalized fluctuation-dissipation theorem. In-gap
collective mode of density fluctuation, known as Bogoliubov-Anderson
phonon, appears as a strong resonance in the plot of DSF as a
function of energy transfer $\omega$ at low momentum transfer $q$.
As the temperature increases, the width of the resonance increases
due to Landau damping. At a large scattering length ($k_F|a_s|>1$),
the resonance may occur at a finite and appreciable value of $q$.
Polarization-selective light scattering may be useful in exciting BA
mode. We have also presented results on single-particle excitation
spectrum when Cooper-pairs are broken ($\omega > 2\Delta$). As the
temperature is decreased, the peak of the spectrum makes a positive
shift due to increase of pair-breaking energy ($2\Delta$). We have
also shown the effect of light polarization on collective as well as
single-particle excitations. Particularly important is the positive
shift of the peak of the single-particle excitation spectrum at
lower temperatures due to circularly polarized light as compered to
that due to unpolarized light (Fig. 4). Furthermore, higher the
momentum transfer, sharper is the single-particle excitation
spectrum. As pointed out earlier \cite{debig}, in stimulated light
scattering, the momentum transfer may be as large as comparable with
the Fermi momentum. Since in polarization-selective stimulated light
scattering, only the atoms with the same spin state are
predominantly scattered \cite{deb}, this kind of light scattering
with large momentum transfer may be useful in detecting the gap
energy and its temperature dependence. In the present paper, our
study is confined to  uniform FS only. Local density approximation
(LDA) may be applied for studying single-particle excitation of a
trapped FS at a large momentum transfer as done in \cite{deb}.
However, for studying collective mode of trapped system, LDA may be
a bad approximation. We hope to address this problem in our future
communication.

\appendix
\section*{Appendix-A}
\setcounter{section}{1}

Here we first calculate the four $T_{ij}$ terms as expressed in Eqs.
(\ref{t11}-\ref{t22}) and then describe the method of calculation of
various integrals. We can express \bearr
 && T_{11} = \beta^{-1}\sum_{n} \frac{1}{i(\omega_n + \nu_m/2) +
 E_{k_+'}}\times \frac{1}{i(\omega_n-\nu_m/2) +
 E_{k_-'}} \nonumber \\
 &=& \frac{\beta^{-1}}{E_{k_+'} - E_{k_-'}  +  i\nu_m} \sum_{n} \left [\frac{1}{i(\omega_n-\nu_m/2) +
 E_{k_-'}} -\frac{1}{i(\omega_n + \nu_m/2) +
 E_{k_+'}}   \right ] \nonumber \\
 &=&  \frac{ n_F(-E_{k_-'})-n_F(-E_{k_+'}) }{E_{k_+'}-E_{k_-'}  + i\nu_m}\eearr
Similarly, \bearr T_{12} = \frac{
n_F(E_{k_-'})-n_F(-E_{k_+'})}{E_{k_+'}+E_{k_-'} + i\nu_m}\eearr
 \bearr T_{21} &=& \frac{
n_F(-E_{k_-'})-n_F(E_{k_+'})}{-E_{k_+'}-E_{k_-'} + i\nu_m}
\nonumber \\
&=& \frac{ n_F(E_{k_+'}) - n_F(-E_{k_-'})}{E_{k_+'}+E_{k_-'} -
i\nu_m}\eearr \bearr T_{22} &=&  \frac{n_F(E_{k_+'})- n_F(E_{k_-'})
}{E_{k_+'}-E_{k_-'}  - i\nu_m}\eearr One can write $n_F(E) = 1/2 -
\tanh(\beta E/2)/2$ and $n_F(-E) = 1/2 + \tanh(\beta E/2)/2$. Using
these relations, we obtain the Eqs. (\ref{t11}-\ref{t22}).

 Let us now consider the integral
$\bar{I}_{i}$ and $J_{i}$. Setting $z=i\nu_m$, we can write \bearr
\tilde{I}_1(q, z) &=& \frac{1}{(2\pi)^2} \int \int \sin\theta
d\theta k^2 dk I_1(k,q,z) \nonumber \\
&=& \frac{1}{4(2\pi)^2} \int k^2 dk \int_{-1}^{1} dy
\mathcal{R}_{+}(k,q,y,z) \eearr \bearr J_2(q, z) =
\frac{1}{4(2\pi)^2}\int k^2 dk \int_{-1}^{1} dy \frac{1}{E_2E_1}
{\mathcal R}_{-}(k,q,y,z)  \eearr where \bearr {\mathcal
R}_{\pm}(k,q,y,z) &=& \frac{T_{+}(k,q)(E_2+E_1)}{(E_2+E_1)^2 - z^2}
\pm \frac{T_{-}(k,q)(E_2-E_1)}{(E_2-E_1)^2 - z^2}, \eearr Here
\bearr T_{\pm} =\tanh(\beta E_2/2)\pm \tanh(\beta E_12), \eearr Here
$y=\cos \theta$. Similarly, we can express \bearr J_1(q, z) =
\frac{1}{4(2\pi)^2}  \int k^2 dk\int_{-1}^{1} dy
\frac{\mathcal{E}}{E_2E_1} {\mathcal R}_{-}(k,q,y,z) \eearr
 \bearr \tilde{I}_2(q, z) =
\frac{1}{4(2\pi)^2}  \int k^2 dk \int_{-1}^{1} dy
\frac{\mathcal{E}^2 - x^2 }{E_2E_1} {\mathcal R}_{-}(k,q,y,z)
 \eearr

In the limit $T\rightarrow 0$, we have $T_{+} = 2$ and $T_{-} =
0$. Now, \bearr \tilde{I}_1 &=& \frac{1}{4(2\pi)^2}  \int k^2 dk \int_{-1}^{1} dy\nonumber \\
&\times& \left
[\frac{T_{+}(E_1+E_2)\{(E_2-E_1)^2-z^2\}+T_{-}(E_2-E_1)\{(E_1+E_2)^2-z^2\}
}{(E_2^2-E_1^2)^2 -
2z^2(E_1^2+E_2^2) + z^4} \right ]\nonumber \\
&=& \frac{1}{4(2\pi)^2}\int k^2 dk  \int_{-1}^{1} dy
\frac{2}{(E_2^2-E_1^2)^2 -
2z^2(E_1^2+E_2^2) + z^4} \nonumber \\
&\times&  \left [E_2\tanh\left(\frac{\beta
E_2}{2}\right)(E_2^2-E_1^2-z^2) - E_1 \tanh\left(\frac{\beta
E_1}{2}\right)(E_2^2-E_1^2+z^2) \right ] \nonumber
 \eearr
 Substituting $E_2^2 - E_1^2 = 4\mathcal{E}x$ and $E_1^2
+ E_2^2 = 2 \mathcal{E}^2 + 2 x^2 + 2\tilde{\Delta}^2$,  the above
equation can then be expressed as \bearr
 \tilde{I}_1 &=& \frac{3\rho_0}{8\epsilon_F \tilde{q}} \int d\mathcal{E} \int_{-\tilde{q} \tilde{k}}^{\tilde{q} \tilde{k}} dx
\frac{1}{z^2 (z^2-4\mathcal{E}^2 - 4\tilde{\Delta}^2)+4(4\mathcal{E}^2-z^2)x^2 } \nonumber \\
&\times& \left [ -z^2 {\mathcal F}_{+} -4\mathcal{E}x{\mathcal
F}_{-}\right ] \nonumber \\
&=& \frac{3\rho_0}{16\epsilon_F \tilde{q}}\int
d\mathcal{E}\frac{1}{z^2(\mathcal{E}^2-\mathcal{Z}_0^2)}\int_{0}^{k
q} dx \Phi(x) \frac{1}{1+\nu^2 x^2} \eearr where \bearr \nu^2
= \frac{z^2-4\mathcal{E}^2}{z^2(\mathcal{E}^2-\mathcal{Z}_0^2)}\\
\nonumber \\
\mathcal{Z}_0^2  = \frac{1}{4}(z^2 - 4 \tilde{\Delta}^2), \\
\nonumber \\
\Phi(x) = z^2 {\mathcal F}_{+} +4\mathcal{E}x{\mathcal F}_{-},
 \eearr    with ${\mathcal F}_{\pm} = E_1
\tanh(\beta E_1/2) \pm E_2 \tanh(\beta E_2/2) $. Here $\rho_0 =
k_F^3/(6\pi^2)$ is the number density of noninteracting Fermi gas of
a single-spin. We further note that $E_1(x) = E_2(-x)$ implying
${\mathcal F}_{\pm}(x) = \pm {\mathcal F}_{\pm} (-x)$. Let us carry
out $x-$integration by parts, assuming $\Phi(x)$  and $(1-\nu^2
x^2)^{-1}$ be the first and second integrand, respectively. Then we
have \bearr I_{\theta} =  \int_{0}^{\tilde{q} \tilde{k}} dx
& &\Phi(x) \frac{1}{1+\nu^2 x^2} = \frac{1}{\nu} \nonumber \\
& \times &\left [ \Phi(\tilde{q} \tilde{k})\tan^{-1}(\nu \tilde{q}
\tilde{k}) - \int_{0}^{\tilde{q} \tilde{k}} dx \Phi'(x)
\tan^{-1}(\nu x) \right ]\eearr

Let us now consider the analytic continuation by taking the limit $z
\rightarrow \tilde{\omega} + i0^{+}$,
 where $\tilde{\omega}$ is the energy transfer. There are two regimes of
 excitations: I. Quasi-particle regime: $\tilde{\omega} > 2\tilde{\Delta}$
 and II. Collective oscillation regime: $\tilde{\omega} < 2\tilde{\Delta}$.
 We first concentrate  on the regime of
quasi-particle excitation, that is, $\tilde{\omega} >
2\tilde{\Delta}$. In this regime, $\mathcal{E}_0 >0$. Let $\phi
=\tan^{-1}(\nu x)$
 and $\mathcal{E}_0 = \sqrt{\tilde{\omega}^2 - 4\tilde{\Delta}^2}/2$. We have three cases for consideration of
 analytic continuation:
(1) For $ |\mathcal{E}| \le \mathcal{E}_0$,  we have
 \bearr \nu
\rightarrow i\nu_1 = i \frac{1}{\tilde{\omega}} \sqrt{
\frac{\tilde{\omega}^2 - 4\mathcal{E}^2}{\mathcal{E}_0^2 -
\mathcal{E}^2 }} \label{cond1} \eearr \bearr \phi &&\rightarrow i
\tanh^{-1}(\nu_1 x), \hspace{0.5cm}  \nu_1x \le 1, \\
\nonumber \\ \phi &&\rightarrow \frac{\pi}{2} + i \tanh^{-1}\left (
\frac{1}{\nu_1 x} \right ), \hspace{0.5cm}  \nu_1x >  1, \eearr (2)
For  $\mathcal{E}_0 < |\mathcal{E}| \le \tilde{\omega}/2$, we have
 \bearr \nu \rightarrow \nu_2 = \frac{1}{\tilde{\omega}}
\sqrt{ \frac{\tilde{\omega}^2 - 4\mathcal{E}^2}{\mathcal{E}^2
-\mathcal{E}_0^2 }}, \\
 \tan\phi \rightarrow \nu_2 x, \eearr (3) For $|\mathcal{E}| >
\tilde{\omega}/2$, we have \bearr
 \nu \rightarrow i\nu_3 = i \frac{1}{\tilde{\omega}} \sqrt{
\frac{4\mathcal{E}^2- \tilde{\omega}^2 }{\mathcal{E}^2
-\mathcal{E}_0^2 }} \eearr
 \bearr \phi &&\rightarrow i
\tanh^{-1}(\nu_3 x), \hspace{0.5cm}  \nu_3x \le 1, \\
\nonumber \\ \phi &&\rightarrow \frac{\pi}{2} + i \tanh^{-1}\left (
\frac{1}{\nu_3 x} \right ), \hspace{0.5cm}  \nu_3x >  1, \eearr

After having done angular integration and analytic continuation, the
energy integration is carried out numerically. Let us next consider
the integral $J_2$. Towards this end, we have \bearr
\frac{\mathcal{R}_{-}}{E_1E_2} &=&2 \frac{z^2\mathcal{J}_{+}(x) +
4\mathcal{E}x\mathcal{J}_{-}(x)}{z^2(\mathcal{E}^2-\mathcal{E}_0^2)(1+\nu^2
x^2)}
 \eearr
where $\mathcal{J}_{\pm}(x) = \tanh[\beta E_2/2]/E_2 \pm \tanh[\beta
E_1/2]/E_1$. Using this, we can write \bearr J_2 =
\frac{3\rho_0}{16\epsilon_F\tilde{q}}\int
d\mathcal{E}\frac{1}{z^2(\mathcal{E}^2-\mathcal{Z}_0^2)}\int_{0}^{
\tilde{q} \tilde{k}} dx \Psi(x) \frac{1}{1+\nu^2 x^2} \eearr where
\bearr \Psi(x) = z^2 {\mathcal J}_{+} +4\mathcal{E}x{\mathcal J}_{-}
 \eearr
Thus $J_2$ integral can be evaluated exactly the same way as done
for
 $\tilde{I}_1$. Similarly, we can find the other two
 integrals
$\tilde{I}_2$ and $J_1$.

\appendix
\section*{Appendix-B}
\setcounter{section}{2}

 We can rewrite the response equation  \bearr
\chi(\mathbf{q},\omega) &=& \int
\frac{d^3k}{(2\pi)^3}\beta^{-1}\sum_n \frac{1}{[(\omega_n+\nu_m)^2 +
E_2^2][\omega_n^2+E_1^2]} \nonumber \\  &\times& \rm{Tr}\left
[\tilde{\gamma}{\mathcal K} \{i(\omega_n -
\nu_m/2)\tau_0+\xi_{k_-}\tau_3 + \tilde{\Delta}\tau_1\}\right
]\eearr where \bearr {\mathcal K} &=& i((\omega_n +
\nu_m/2))\sum_{j=0}^{3}\Gamma^{(j)}\tau_j + \xi_{k_+}\left \{
\Gamma^{(0)}\tau_3 + \Gamma^{(3)}\tau_0 +
i\sum_{j=1,2}\epsilon_{3j}\Gamma^{(j)}\tau_{3-j}\right \} \nonumber
\\ &+&  \tilde{\Delta} \left \{ \Gamma^{(0)}\tau_1 + \Gamma^{(1)}\tau_0 +
i\sum_{j=2,3} \Gamma^{(j)} \epsilon_{1j}\tau_{n_j} \right \}\eearr
where $n_j = |1+\epsilon_{1j}j|$ and
$\epsilon_{ij}=-\epsilon_{ji}=1$ if $(i,j)$ is (1,2) or (2,3) or
(3,1).  Now, \bearr {\mathcal K} &\times& \{i(\omega_n -
\nu_m/2)\tau_0+\xi_{k_-}\tau_3 + \tilde{\Delta}\tau_1\} = i(\omega_n
- \nu_m/2)\mathcal{K} \nonumber \\
&+&  \xi_{k_-}i((\omega_n + \nu_m/2)) \left \{ \Gamma^{(0)}\tau_3 +
\Gamma^{(3)}\tau_0 +
i\sum_{j=1,2}\epsilon_{j3}\Gamma^{(j)}\tau_{3-j}\right \} \nonumber
\\ &+&
 \xi_{k_+}\xi_{k_-} \left \{ \Gamma^{(0)}\tau_0 +
\Gamma^{(3)}\tau_3 -
 \sum_{j=1,2} \Gamma^{(j)}\tau_j \right
\} \nonumber \\ &+& \tilde{\Delta} \xi_{k_-} \left \{
-i\Gamma^{(0)}\tau_2 + \Gamma^{(1)}\tau_3 + i \Gamma^{(2)}\tau_0 +
\Gamma^{(3)}\tau_1
\right \} \nonumber \\
&+& i\tilde{\Delta} (\omega_n + \nu_m/2) \left \{ \Gamma^{(0)}\tau_1
+ \Gamma^{(1)}\tau_0 + i\sum_{j=2,3} \Gamma^{(j)}
\epsilon_{j1}\tau_{n_j} \right \} + \tilde{\Delta} \xi_{k_+} \left
\{
i\Gamma^{(0)}\tau_2 \right.  \nonumber \\
 &+& \left.
\Gamma^{(1)}\tau_3 - i \Gamma^{(2)}\tau_0 + \Gamma^{(3)}\tau_1
\right \} + \tilde{\Delta}^2 \left \{ \Gamma^{(0)}\tau_0 +
\Gamma^{(1)}\tau_1 - \sum_{j=1,3} \Gamma^{(j)}\tau_j \right \}
\eearr So, \bearr &\rm{Tr}&[\tilde{\gamma} \mathcal{K}(i(\omega_n -
\nu_m/2)\tau_0+\xi_{k_-}\tau_3 + \tilde{\Delta}\tau_1)] \nonumber
\\
&=&
-2(\omega_n-\nu_m/2)(\omega_n + \nu_m/2) \sum_{j=0,3}\gamma_j\Gamma^{(j)}\nonumber \\
&+& 2i(\omega_n - \nu_m/2)\xi_{k_+}[\gamma_0\Gamma^{(3)} + \gamma_3
\Gamma^{(0)}] + 2i(\omega_n - \nu_m/2)\tilde{\Delta} [ \gamma_0
\Gamma^{(1)} + i\gamma_3 \Gamma^{(2)} ] \nonumber \\ &+&
2\xi_{k_-}i((\omega_n + \nu_m/2))\left
\{ \gamma_3\Gamma^{(0)} + \gamma_0\Gamma^{(3)}\right \} \nonumber \\
&+& 2\xi_{k_+}\xi_{k_-} \left \{ \Gamma^{(0)}\gamma_0 +
\Gamma^{(3)}\gamma_3  \right \} + 2\tilde{\Delta} \xi_{k_-} \left \{
 \Gamma^{(1)}\gamma_3 + i \Gamma^{(2)}\gamma_0
\right \} \nonumber \\
&+& 2i\tilde{\Delta} (\omega_n + \nu_m/2) \left \{
\Gamma^{(1)}\gamma_0
- i \Gamma^{(2)}\gamma_3 \right \}  \nonumber \\
&+& 2\tilde{\Delta} \xi_{k_+} \left \{ \Gamma^{(1)}\gamma_3 - i
\Gamma^{(2)}\gamma_0   \right \} + 2\tilde{\Delta}^2 \left \{
\Gamma^{(0)}\gamma_0   -  \Gamma^{(3)}\tau_3 \right \} \eearr

\appendix
\section*{Appendix-C}
\setcounter{section}{3}

Here we  turn our attention to collective oscillation regime where
$\mathcal{E}_0^2 < 0$. We do not make any attempt to evaluate
analytically the collective mode for any interaction strength
($k_F|a_s|$) at finite temperature. However, it is possible to
calculate analytically BA mode energy in the weak-coupling limit
$k_F|a_s|\rightarrow 0$ at zero temperature. Towards this end, let
us first calculate  $D_+(z\rightarrow \omega + i 0^+)$. Restricting
$x < 1$, we expand the functions $E_{\pm} \equiv E_{2,1}(x)$, and
$1/E_{\pm}(x)$ up to second order in $x$ \bearr E_{\pm} \simeq E +
\frac{x^2 \pm 2\mathcal{E}x}{2E} - \frac{\mathcal{E}^2x^2}{2E^3}
\eearr  \bearr \frac{1}{E_{\pm}} \simeq \frac{1}{E}\left[1 -
\frac{x^2 \pm 2\mathcal{E}x}{2E^2} +
\frac{3\mathcal{E}^2x^2}{2E^4}\right ] \eearr

At zero temperature,  we obtain \bearr \Phi(x\simeq 0) &\simeq&
  \tilde{\omega}^2(2E + \frac{x^2}{E}) -
8\mathcal{E}^2\frac{x^2}{E} -
\frac{\tilde{\omega}^2\mathcal{E}^2x^2}{E^3}, \eearr \bearr
\Psi(x\simeq 0) \simeq  \frac{1}{E} \left [ \tilde{\omega}^2(2 -
\frac{x^2}{E^2}) - 8\mathcal{E}^2 \frac{x^2}{E^2}  +
\frac{3\tilde{\omega}^2\mathcal{E}^2x^2}{E^4} \right ] \eearr We
can write $ \tilde{I}_1 + \tilde{I}_2 + \Delta^2 J_2 \simeq I_< +
I_>$, where
 \bearr I_<  & = & \frac{3\rho_0}{4\epsilon_F}\int_{|\mathcal{E}| \le
\tilde{\omega}/2} d\mathcal{E} \left [
\frac{E}{(\mathcal{E}^2-\mathcal{E}_0^2)}\frac{\tan^{-1}(\nu_2
\tilde{q}
\tilde{k})}{\nu_2 \tilde{q}} \right. \nonumber \\
&-& \left.
\frac{1}{\tilde{\omega}^2(\mathcal{E}^2-\mathcal{E}_0^2)}\left (
\frac{\tilde{\omega}^2+8\mathcal{E}^2}{2E} - \frac{2\tilde{\omega}^2
\mathcal{E}^2}{E^3}\right ) \left ( \frac{\tilde{k}}{\nu_2^2} -
\frac{\tan^{-1}(\nu_2 \tilde{q} \tilde{k})}{\nu_2^3\tilde{q}} \right
) \right ] \eearr  Here $I_<$ is real, but $I_>$ has both real and
imaginary parts. We now calculate \bearr \rm{Re}[ I_{>}] &=&
\frac{3\rho_0}{4\epsilon_F }\int_{\mathcal{E}_{c}>|\mathcal{E}|
> \tilde{\omega}/2} d\mathcal{E}  \left [
\frac{E}{(\mathcal{E}^2-\mathcal{E}_0^2)}\frac{\tanh^{-1}(\nu_3
\tilde{q}
\tilde{k})}{\nu_3 \tilde{q}} \right. \nonumber \\
&-& \left.
\frac{1}{\tilde{\omega}^2(\mathcal{E}^2-\mathcal{E}_0^2)}\left (
\frac{\tilde{\omega}^2+8\mathcal{E}^2}{2E} - \frac{2\tilde{\omega}^2
\mathcal{E}^2}{E^3}\right ) \left ( - \frac{k}{\nu_3^2} +
\frac{\tanh^{-1}(\nu_3 \tilde{q} \tilde{k})}{\nu_3^3\tilde{q}}
\right ) \right ] \nonumber \\
&+& \frac{3\rho_0}{4\epsilon_F}\int_{ |\mathcal{E}|
>\mathcal{E}_{c}} d\mathcal{E}
 \left [ \frac{E}{(\mathcal{E}^2-\mathcal{E}_0^2)}\frac{\tanh^{-1}[1/(\nu_3
\tilde{q}
\tilde{k})]}{\nu_3 \tilde{q}} \right. \nonumber \\
&-& \left.
\frac{\tilde{\omega}^2+8\mathcal{E}^2}{2E\tilde{\omega}^2(\mathcal{E}^2-\mathcal{E}_0^2)}
\left ( - \frac{\tilde{k}}{\nu_3^2} + \frac{\tanh^{-1}[1/(\nu_3
\tilde{q} \tilde{k})]}{\nu_3^3\tilde{q}} \right ) \right] \eearr
where $\mathcal{E}_c$ is  the energy $\mathcal{E}$ which is the
root of the equation $\nu_3 \tilde{q}\tilde{k} = 1$. In the
weak-coupling limit, we can carry out the energy integration over
the Fermi surface. In that case, the integration over the energy
$|\mathcal{E}|> \mathcal{E}_c$ is negligible and $\mathcal{E}_c$
may be assumed to tend to $\infty$. Under this condition, putting
$k \simeq k_{\mu} \simeq k_F$, in the zero temperature limit,
carrying out the Fermi surface integral, expanding the terms like
$\tan^{-1}(\tilde{q} \tilde{k}) $ and $\tanh^{-1}(\tilde{q}
\tilde{k})$ up to the second order in $\tilde{q}$,  we obtain
\bearr \rm{Re}[D_{+}] &\simeq& \frac{3\rho_0
\tilde{k}_{\mu}}{4\epsilon_F} \int d \mathcal{E} \left [ \left (
\frac{E}{(\mathcal{E}^2-\mathcal{E}_0^2)} - \frac{1}{E} \right )
- \frac{(\tilde{q} \tilde{k}_{\mu})^2}{3}\left ( \frac{1}{2E(\mathcal{E}^2-\mathcal{E}_0^2)} \right ) \right. \nonumber \\
&-& \left. \frac{(\tilde{q} \tilde{k}_{\mu})^2}{3} \left ( \frac{E
\nu_2^2 }{(\mathcal{E}^2-\mathcal{E}_0^2)} + \frac{4 \mathcal{E}^2}{
\tilde{\omega}^2 E (\mathcal{E}^2-\mathcal{E}_0^2)} - \frac{2
\mathcal{E}^2}{ E^3 (\mathcal{E}^2-\mathcal{E}_0^2)} \right) \right]
\\
&=& \frac{3\rho_0 \tilde{k}_{\mu}}{4\epsilon_F} \left [ \left (
\frac{1}{2} -  \frac{(q k_{\mu})^2}{3\tilde{\omega}^2}\right )\left
( \frac{\bar{\omega}}{\sqrt{1-\bar{\omega}^2}} \right )\left (  \pi
- 2 \sin^{-1}(\sqrt{1-\bar{\omega}^2}) \right )
\right. \nonumber \\
&-& \left. \frac{(\tilde{q} \tilde{k}_{\mu})^2 }{3\tilde{\omega}^3
\Delta } \left ( \frac{ 2\tilde{\omega} \Delta
\sqrt{1-\bar{\omega}^2} - 4 \Delta^2 (1 -
2\bar{\omega}^2)\sin^{-1}(\bar{\omega})}{(1-\bar{\omega}^2)^{3/2}}
\right ) \right. \nonumber \\
&+&  \left. \frac{(\tilde{q} \tilde{k}_{\mu})^2}{3} \left
(\frac{4\Delta}{\tilde{\omega}^3} \right) \left (\bar{\omega} -
\sqrt{1-\bar{\omega}^2}\sin^{-1}\bar{\omega}\right)\right ]
\label{fsint} \eearr where
$\bar{\omega}=\tilde{\omega}/(2\tilde{\Delta})$. For $\bar{\omega}
<\!<1$, we can expand \bearr \sin^{-1}(\sqrt{1-\bar{\omega}^2})
&\simeq& \sin^{-1}(1-\frac{1}{2}\bar{\omega}^2) = \frac{\pi}{2} -
\bar{\omega} - \frac{1}{12}\bar{\omega}^3 + \cdots \eearr Similarly,
expanding $\sin^{-1}(\bar{\omega})$  and retaining the terms lower
order in $\bar{\omega}$,  we can approximate the square bracketed
part of Eq. (\ref{fsint}) as \bearr &&
\frac{1}{\sqrt{1-\bar{\omega}^2}}\left [ \bar{\omega}^2 +
\frac{1}{12} \bar{\omega}^4 - \frac{(\tilde{q}
\tilde{k})^2}{3(2\Delta)^2}\left ( 2 + \frac{1}{6}\bar{\omega^2}
\right )\right ] \nonumber \\ &-& \frac{8(\tilde{q}
\tilde{k}_{\mu})^2}{9(2\Delta)^2(1-\bar{\omega}^2)^{3/2}} +
\frac{2(\tilde{q} \tilde{k}_{\mu})^2}{9(2\Delta)^2}  \eearr
Furthermore, keeping the terms up to second order in $\bar{\omega}$
and neglecting the product terms like $(\tilde{q} \tilde{k}_{\mu})^2
\bar{\omega}^2$, the above expression reduces to \bearr
\bar{\omega}^2 - \frac{4}{3} \left (\frac{\tilde{q}
\tilde{k}_{\mu}}{2\tilde{\Delta}}\right)^2 \eearr Equating this to
zero, we find the root $\tilde{\omega} = \frac{2}{\sqrt{3}}\tilde{q}
\tilde{k}_{\mu}$ which implies \bearr \omega = \frac{1}{\sqrt{3}}
p_q v_{\mu} \eearr where $p_q = \hbar q $ and $v_{\mu} = \hbar
k_{\mu}/m$.


\section*{References}
\begin{thebibliography}{10}

\bibitem{bec}  Anderson M.,  Ensher J. R.,  Matthews M. R.,
 Wieman C. E., and Cornell  E. A. 1995 {\it Science} {\bf 269} 198;
  Bradley C.
C.,  Sackett C. A., Tollett J. J. , and  Hulet R. G. 1995 {\it Phys.
Rev. Lett.} {\bf 75}  1687 ;  Davis K. B. ,  Mewes M. O., Andrews M.
R., van Druten N. J.,  Durfee D. S., Kurn D. M. , and Ketterle W.
1995 {\it Phys. Rev. Lett.} {\bf 75}  3969

\bibitem{hulet} A. G. Truscott, K. E. Strecker, W. I. McAlexander, G.B.
Patridge, and R. G. Hulet 2001 {\it Science} {\bf 291} 2570

\bibitem{solomon} F. Schreck {\it et al.}  2001 {\it Phys. Rev. Lett.} {\bf 87}
080403;T. Bourdel {\it et al.} 2003 {\it ibid.} {\bf 91} 020402

\bibitem{thomas} S. R. Granade, M. E. Gehm, K. M. O'Hara, and J. E.
Thomas  2002 Phys.Rev.Lett. {\bf 88}, 120405;  O'Hara  {\it et al.}
2002  {\it Science} {\bf 298}  2179

\bibitem{mit} Z. Hadzibabic {\it et al.},
 {\it Phys. Rev. Lett.} {\bf 88}, 160401 (2002).

\bibitem{italy} G. Roati, F. Riboli, G. Modungo, and M. Inguscio,
{\it Phys. Rev. Lett.} {\bf 89}, 150403 (2002).

\bibitem{grimm} Jochim S. {\it et al.} 2003  {\it Science} {\bf 302}
 2101

\bibitem{expt}  Modugno G. {\it et al.} 2002 {\it Science} {\bf 297}
  2240;  Strecker K. E. {\it et al.} 2003 {\it Phys. Rev. Lett.}
{\bf 91} 080406;  Cubizolles J. {\it et al.} 2003 {\it Phys. Rev.
Lett.} {\bf 91}  240401

\bibitem{jin} B. DeMacro and D. S. Jin 1999 {\it Science} {\bf 285} 1703

\bibitem{ketterle} Zwierlein M. W., Abo-Shaeer J. R., Schirotzek A.,
Schunck C. H., Ketterle W. 2005 {\it Nature} {\bf 435}  1047

\bibitem{gap1} Cin C. {\it et al.} 2004 {\it Science} {\bf 305}
1128

\bibitem{gap2} Greiner M., Regal C. A., and Jin
D. S. 2005 {\it Phys. Rev. Lett.} {\bf 94} 070403

\bibitem{theogap} Kinnunen J., Rodriguez M., and T\"{o}rm\"{a} P
 2004
{\it Science} {\bf 305} 1131

\bibitem{zoller} T\"{o}rm\"{a} P. and  Zoller P. 2000 {\it Phys. Rev. Lett.} {\bf 85}  487;
 Bruun G. M. {\it et al.} 2001 {\it Phys. Rev. A} {\bf 64}
 033609;  Kinnunen J., Rodriguez M., and T\"{o}rm\"{a} P 2004 Phys. Rev. Lett. {\bf
 92} 230403; Bruun G. M. and Baym G. 2004 {\it Phys. Rev. Lett.} {\bf 93}
 150403; B$\ddot{u}$chler H. P.,  Zoller P., Zwerger W. 2004 {\it Phys. Rev. Lett.}, {\bf
 93}  080401

\bibitem{huletp}  Zhang W.,  Sackett C. A. and  Hulet R. G. 1999 {\it Phys. Rev. A} {\bf 60} 504;
 Ruostekoski J. 1999 {\it Phys. Rev. A}  {\bf 60}  1775;
 Rodriguez M.
and T\"{o}rm\"{a} P. 2002 {\it Phys. Rev. A} {\bf 66} 033601.

\bibitem{duke} Kinast, J. {\it et al.} 2004  {\it Phys. Rev. Lett.} {\bf 92} 150402

\bibitem{innsbruck} Bartenstein M. {\it et al.} 2004 {\it Phys. Rev. Lett.} {\bf 92}
203201

\bibitem{stringari} Stringari S. 2004 {\it Europhys. Lett.} {\bf 65}
749

\bibitem{nozrink}  Nozi$\acute{e}$res  P. and  Schmitt-Rink S. 1985 {\it J.
Low. Temp. Phys.} {\bf 59}  195

\bibitem{randeria}  Sa de Melo C.A.R.,
Randeria M. and  Engelbrecht J.R. 1993 {\it Phys. Rev. Lett.} {\bf
71}  3202; Engelbrecht J.R., Randeria M. and Sa de Melo C.A.R. 1997
{\it Phys. Rev. B} {\bf 55} 15153

\bibitem{crossover}  Holland M.,  Kokkelmans S. J. J. M. F.,
Chiofalo M. L. and  Wasler R. 2001 {\it Phys. Rev. Lett.} {\bf 87}
 120406;  Timmermans E. {\it et al.} 2001 {\it Phys. Lett A} {\bf 285}
  228;  Ohashi Y. and  Griffin  A. 2002 {\it Phys. Rev. Lett.} {\bf 89}
  130402;  Hofstetter W. {\it et al. } 2002 {\it Phys. Rev. Lett.} {\bf
89} 220407

\bibitem{molecules}  Greiner M.,  Regal C. A. and  Jin D. S. 2003 {\it Nature} {\bf 426}
 537; Jochim S. {\it et al.}  2003 {\it Science} {\bf 302}  2101;
Zwierlein M. W. {\it et al.} 2003 {\it Phys. Rev. Lett.} {\bf 91}
 250401

 \bibitem{theory}  Falco G. M. and  Stoof H. T. C. 2004 {\it Phys. Rev. Lett.} {\bf 92}   130401;
 Carr L. D.,  Shlyapnikov G. V. and  Castin Y. 2004 {\it Phys. Rev.
 Lett.}
{\bf 92} 150404;  Heiselberg H. 2003 {\it Phys. Rev. A} {\bf 68}
 053616,  Perali A.,  Pieri P. and  Strinati G. C. 2003 {\it
Phys. Rev. A}, {\bf 68}  031601;  Perali A.,  Pieri P., Pisani L.
and  Strinati G. C. 2004 {\it Phys. Rev. Lett.} {\bf 92} 220404;
Ohashi Y. and Griffin A. 2005 {\it Phys. Rev. A} {\bf 72} 013601;
{\it ibid} 063606

 \bibitem{huletim} Guthrie B. {\it et al.} 2006 {\it Science} {\bf 311}
 503

 \bibitem{ketterleim} Zwierlein M. W. {\it et al.} 2006 {\it Science} {\bf
 311} 492

 \bibitem{liu}  W. V. Liu and F. Wilczek 2003
{\it Phys.Rev.Lett.} {\bf 90} 047002

\bibitem{debig}  Deb B. {\it et al.}  2004 {\it Phys. Rev. A}  {\bf 70}  011604

\bibitem{deb} Deb B. 2006 {\it J.
Phys. B: At. Mol. \& Opt. Phys.} {\bf 39} 529

\bibitem{bamode} Bogoliubov N. N. 1958 {\it Nuovo Cimento} {\bf 7} 6;
Bogoliubov N. N., Tolmachev V. V., and Shirkov D. V. 1959 {\it A New
Method in the Theory of Superconductivity} (Consultants Bureau, NY);
Anderson P. W. 1958 {\it Phys. Rev.}  {\bf 112} 1900

\bibitem{mottelson} Bruun G. M. and Mottelson B. R. 2001 {\it Phys. Rev.
Lett.} {\bf 87}  270403; Ohashi Y. and Griffin A. 2003 {\it Phys.
Rev. A} {\bf 67} 063612;   Minguzzi A.,  Ferrari G. and  Castin Y.
2001 {\it Eur. Phys. J. D.} {\bf 17} 49; Heiselberg H. 2006 {\it
Phys. Rev. A} {\bf 73} 013607

\bibitem{schrieffer}  Schrieffer J. R. 1964 {\it Theory of
Superconductivity} ( W. A. Benjamin)

\end{thebibliography}
 \end{document}